\documentclass[pra,aps,groupaddress,showpacs,twocolumn]{revtex4}
\usepackage{epsfig,amsmath,graphicx,amssymb,overpic}
\usepackage{bm}
\def\be{\begin{equation}}
\def\ee{\end{equation}}
\def\bee{\begin{eqnarray}}
\def\ene{\end{eqnarray}}
\def\bes{\begin{subequations}}
\def\ees{\end{subequations}}

\newcommand{\PT}{\mathcal{PT}}

\begin{document}
\title{Stable parity-time-symmetric nonlinear modes and excitations \\ in a derivative nonlinear Schr\"odinger equation}
\author{Yong Chen}
\author{Zhenya Yan\footnote{Corresponding author,\,email adress: zyyan@mmrc.iss.ac.cn}}
\affiliation{Key Laboratory of Mathematics Mechanization, Institute of Systems
Science, AMSS,  Chinese Academy of Sciences, Beijing 100190, China\\
University of Chinese Academy of Sciences, Beijing 100049, China\vspace{0.1in}}

\date{\vspace{0.1in}21 May 2016, Phys. Rev. E  {\bf 95}, 012205 (2017)}


\begin{abstract}
The effect of derivative nonlinearity and parity-time- ($\PT$-) symmetric potentials on the wave propagation dynamics is investigated in the derivative nonlinear Schr\"odinger equation, where the physically interesting Scarff-II and hamonic-Hermite-Gaussian potentials are chosen. We study numerically the regions of unbroken/broken linear $\PT$-symmetric phases and find some stable bright solitons of this model in a wide range of potential parameters even though the corresponding linear $\PT$-symmetric phases are broken. The semi-elastic interactions between exact bright solitons and exotic incident waves are illustrated such that we find that exact nonlinear modes almost keep their shapes after interactions even if the exotic incident waves have evidently been changed. Moreover, we exert the adiabatic switching on $\PT$-symmetric potential parameters such that a stable nonlinear mode with the unbroken linear $\PT$-symmetric phase can be excited to another stable nonlinear mode belonging to the broken linear $\PT$-symmetric phase.
\end{abstract}

\maketitle


\section{Introduction}

The derivative nonlinear Schr\"odinger (DNLS) equation
\bee\label{dnls0}
i\psi_t+\psi_{xx}+ig(|\psi|^2\psi)_x=0, \quad g>0
\ene
where $g$ represents its relative magnitude (the space-reflection transformation $x\to -x$ can make $g<0$) and the derivative nonlinearity term is also called the nonlinear dispersion term~\cite{keer}. In fact, Eq.~(\ref{dnls0}) has a close relation with the modified nonlinear Schr\"odinger (MNLS) equation~\cite{phap4, phap5}
\be\label{dnls0g}
iq_{\xi}+\alpha q_{\tau\tau}+\lambda|q|^2q+i\gamma(|q|^2q)_{\tau}=0,
\ee
where $\alpha$ denotes the group velocity dispersion coefficient, the Kerr nonlinear coefficient $\lambda$ and derivative nonlinear coefficient $\gamma$ both depend on nonlinear refractive index $n_2$. Eq.~(\ref{dnls0g}) can be transformed into Eq.~(\ref{dnls0}) by using the similarity transformation~\cite{cl04} $q(\tau, \xi)=\psi(x, t)e^{i(kx+k^2t)}$ with $x=\frac{\gamma}{\alpha g}\tau-\frac{2\lambda}{g}\xi$,\, $t=\frac{\gamma^2}{\alpha g^2}\xi$, and $k=\frac{\alpha g\lambda}{\gamma^2}$.
Eq.~(\ref{dnls0}) (or the similarity Eq.~(\ref{dnls0g})) can be used to describe many nonlinear wave phenomena in some physical applications such as the propagations of small-amplitude nonlinear Alfv\'en waves in a low-$\beta$ plasma~\cite{dnlsp}, large-amplitude magnetohydrodynamic waves propagating in an arbitrary direction with respect to the magnetic field in a high-$\beta$ plasma~\cite{dnls-02}, the filamentation of lower-hybrid waves~\cite{dnls-77}, and the sub-picosecond or femtosecond pulses in single-mode optical fiber~\cite{phap4, phap5}. Eq.~(\ref{dnls0}) can be solved using the inverse scattering method~\cite{kn78}. Moreover, its some modified versions (e.g., Eq.~(\ref{dnls0g}) had also been studied such as the Chen-Li-Liu equation~\cite{cll} and the modified NLS equation~\cite{cll,cons79,cl04}.

The NLS equation describing light propagation in optics~\cite{opso} with real external potentials or/and gain-and-loss distributions has been investigated~\cite{simi0, yanpla10, yanpre12, 1dnlsv, 1dnlsv2, yanpre15, mu08, vvk12, pt1, pt2, pt3, exp9, exp10, exp11, exp12, exp13, exp14} since the refractive index of the optical waveguide can be complex~\cite{complex1, complex2}. It is surprising to find that if the complex refractive index satisfies the property of the parity-time ($\PT$) symmetry~\cite{bender}, that is, if the real and imaginary parts of the refractive index are the even and odd functions of spatial position, respectively, then the propagation constant of the light can still be in all-real spectrum range, hence admitting stationary beam transmission~\cite{real1, real2, real3, real4}. Moreover, the complex $\PT$-symmetric potentials can also support continuous families of stable solitons~\cite{yanpre15, mu08, vvk12, pt1,pt2, pt3, exp9, exp10, exp11, exp12, exp13, exp14} even if the solitons appear in the range of the broken linear $\PT$-symmetric phases (see, e.g., Ref.~\cite{yanpre15}). More recently, the stable nonlinear modes were found in the third-order NLS equation with $\PT$-symmetric potentials~\cite{yan16}. Other interesting $\PT$-symmetric phenomena or properties can be found in the
relevant experimental studies~\cite{real2, real3, exp3, exp4}.

It is still a significant subject to study whether stable nonlinear modes exist in other models with $\PT$-symmetric potentials. To the best of our knowledge, soliton dynamics of the DNLS equation (\ref{dnls0}) (it can be regarded as the extension of the NLS equation) in the $\PT$-symmetric potentials was not studied before. Our main goal in this paper is to find stable solitons and study their dynamical behaviors of the DNLS equation (\ref{dnls0}) in two kinds of physically interesting $\PT$-symmetric potentials (i.e., $\PT$-symmetric Scarff-II and hamonic-Hermite-Gaussian potentials).

 The rest of this paper is arranged as follows. We firstly present the broken/unbroken regions of the linear spectral problem with $\PT$-symmetric potentials. And then we analysis the effect of the $\PT$-symmetric potentials and derivative nonlinearity on the stability, wave propagations, interactions, transverse power-flow density of solitons in detail. Finally, based on the adiabatic change technique we also perform some types of stable excitations belonging to the broken linear $\PT$-symmetric phases from the nonlinear modes.

\section{Nonlinear physical model with $\PT$-symmetric potentials}

\subsection{The nonlinear model}

We begin our investigation by considering the wave propagations in the derivative nonlinearity and $\PT$-symmetric potentials, which can be modelled by the following normalized derivative nonlinear Schr\"odinger-like equation with $\PT$-symmetric potentials
\be\label{dnls}
i\psi_t+\psi_{xx}-[V(x)+iW(x)]\psi+ig(|\psi|^2\psi)_x=0,
\ee
where $\psi=\psi(x,t)$ is a complex wave function of $x,t$, which is proportional to the electric field envelope, $t$ denotes the scaled propagation time or distance, $x$ represents the normalized transverse coordinate, and $g$ is a positive nonlinear coefficient (without loss of generality we can choose $g=1$). When we make the transformation $t\rightarrow z$ (propagation distance) and $x\rightarrow t$ (propagation time), the above-mentioned model may be used to describe the evolution of pulses inside a single-mode fiber~\cite{keer,keer2}. The $\PT$-symmetric potential $V(x)+iW(x)$ requires that its real and imaginary components satisfy $V(-x)=V(x)$ and $W(-x)=-W(x)$ describing the real-valued external potential and gain-and-loss distribution, respectively. It is easy to show that Eq.~(\ref{dnls}) is invariant under the $\PT$-symmetric transformation if the complex potential $[V(x)+iW(x)]$ is $\PT$-symmetric, where ${\mathcal P}$ and ${\mathcal T}$ operators are defined by ${\mathcal P}:\, x\to -x$; ${\mathcal T}:\, i\to-i,\, t\to-t$. Eq.~(\ref{dnls}) can be rewritten as the form $\psi_t=-\frac{\partial}{\partial x}\frac{\delta\mathcal{H}}{\delta\psi^{*}}$ with the Hamiltonian
$\nonumber\mathcal{H}=\int_{-\infty}^{+\infty}\{-i\psi_x\psi^{*}+\psi^{*}\int_0^x[iV(x)-W(x)]\psi dx+\frac{g}2|\psi|^4\} dx$,
where the asterisk stands for the complex conjugate. The power and quasi-power of Eq.~(\ref{dnls}) are given by $P(t)=\int_{-\infty}^{+\infty}|\psi(x,t)|^2dx$ and $Q(t)=\int_{-\infty}^{+\infty}\psi(x,t)\psi^{*}(-x,t)dx$, respectively. One can immediately obtain that $P_t=2\int_{-\infty}^{+\infty}W(x)|\psi(x,t)|^2dx$ and $Q_t=-\int_{-\infty}^{+\infty}g\psi(x,t)\psi^{*}(-x,t)[(|\psi(x,t)|^2)_x-(|\psi(-x,t)|^2)_x+\psi^{*}(x,t)\psi_x(x,t)-\psi(-x,t)\psi^*_x(-x,t)]dx$.

\subsection{General theory}

The stationary solutions of Eq.~(\ref{dnls}) are considered in the form $\psi(x,t)=\phi(x) e^{i\mu t}$, where $\mu$ is the real propagation constant and the nonlinear localized eigenmode ($\lim_{|x|\rightarrow\infty}\phi(x)=0$) satisfies
\bee\label{ode}
\phi_{xx}-[V(x)+iW(x)]\phi+ig(|\phi|^2\phi)_x=\mu\phi(x).
\ene
For Eq.~(\ref{ode}) with some functions $V(x)$ and $W(x)$, there exist two cases for the study of solutions of Eq.~(\ref{ode}): (i) if $\phi(x)$ is a real-valued function, then we have the solution of Eq.~(\ref{ode})
\bee \label{solua}
 \phi^2(x)=\frac{2}{3g}\partial_x^{-1}W(x),
\ene
with the condition linking the potential and gain-and-loss distribution being
\bee
 W^2(x)\!-\!2W_x(x)\partial_x^{-1}\!W(x)\!+\!4[V(x)\!+\!\mu](\partial_x^{-1}\!W(x))^2\!=\!0, \,\,
\ene
where $\partial_x^{-1}W(x)=\int_0^xW(s)ds$.

(ii) if the function $\phi(x)$ is complex in the form
\bee \label{solub}
 \phi(x)=\rho(x)\exp\left[i\int^x_0v(s)ds\right],
\ene
where $\rho(x)$ is the real amplitude, and the real function $v(x)$ is the hydrodynamic velocity, then we substitute Eq.~(\ref{solub}) into Eq.~(\ref{ode}) to yield the relations linking the hydrodynamic velocity
\bee\label{ode1}
 v(x)=\rho^{-2}(x)\int^x_0W(s)\rho^2(s)ds-\frac{3g}{4}\rho^2(x),
\ene
and the amplitude satisfying the second-order ordinary differential equation with varying coefficients
\bee \label{ode2}
 \rho_{xx}(x)=[V(x)+v^{2}(x)+\mu]\rho(x)+gv(x)\rho^3(x).
\ene

In order to further study the linear stability of such nonlinear localized mode $\psi(x,t)=\phi(x) e^{i\mu t}$, we consider the perturbed solutions of Eq.~(\ref{dnls}) as follow
\be\label{pert}
\psi(x,t)\!=\!\left\{\phi(x)\!+\!\epsilon\!\left[F(x)e^{i\delta t}\!+\! G^*(x)e^{-i\delta^* t}\right]\right\}e^{i\mu t},
\ee
where $\epsilon\ll 1$, $F(x)$ and $G(x)$ are the perturbation eigenfunctions of the linearized eigenvalue problem and $\delta$ measures the growth rate of the perturbation instability. Substituting Eq.~(\ref{pert}) into Eq.~(\ref{dnls}) and linearizing with respect to $\epsilon$, we obtain the following linear eigenvalue problem for the perturbation modes
\bee \label{st}
\left(\begin{array}{cc}   \hat{L}_1 & \hat{L}_2 \vspace{0.05in}\\   -\hat{L}_2^* & -\hat{L}_1^* \\  \end{array}\right)
\left(  \begin{array}{c}    F(x) \vspace{0.05in} \\    G(x) \\  \end{array} \right)
=\delta \left(  \begin{array}{c}   F(x) \vspace{0.05in}\\    G(x) \\  \end{array}\right),
\label{stable}
\ene
where $\hat{L}_1=\partial^2_x+2ig[|\phi|^2\partial_x+(|\phi|^2)_x]-[V(x)+iW(x)]-\mu$ and $\hat{L}_2=ig[\phi^2 \partial_x+(\phi^2)_x]$. Obviously,
the $\PT$-symmetric nonlinear modes are linearly stable if $\delta$ is purely real, otherwise they are linearly
unstable.

In what follows we study Eqs.~(\ref{dnls}) and (\ref{ode}) analytically and numerically in detail for two distinct physically interesting $\PT$-symmetric potentials.

 \begin{figure*}[!t]
 	\begin{center}
 	\vspace{0.05in}
 	\hspace{-0.05in}{\scalebox{0.7}[0.7]{\includegraphics{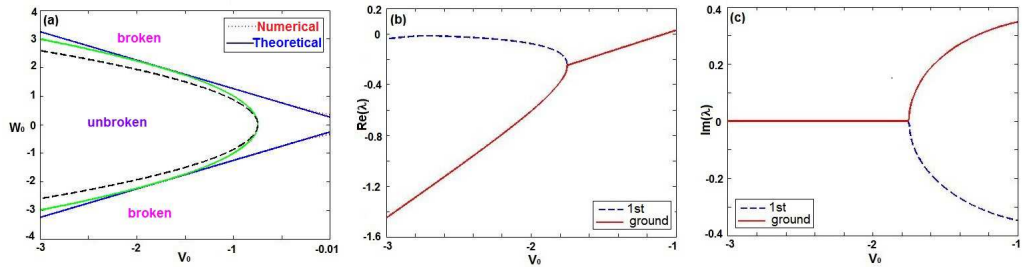}}}
 	\end{center}
 	\vspace{-0.15in} \caption{\small (color online). (a) The unbroken (broken) $\PT$-symmetric phase is in the domain inside (outside) two phase breaking lines (the blue solid theoretically and the red dotted numerically coincide by and large) for the linear operator $L$ in Eq.~(\ref{lp}) with $\PT$-symmetric Scarff-II potentials (\ref{ps}), where the solid parabola is $W_0^2+4V_0+3=0$, whose tangent points with the two phase breaking lines above are $(V_0, W_0)=(-1.75,\pm2)$, and the dashed parabola is $4W_0^2+12V_0+9=0$, which is tangent with the solid parabola at $(V_0, W_0)=(-0.75,0)$. (b) Real and (c) imaginary parts of the eigenvalues $\lambda$ of the linear problem (\ref{lp}) with $\PT$ symmetric potential (\ref{ps}) as a function of $V_0$ at $W_0=2$. The phase transition threshold is approximately $V_0=-1.75$, which coincides exactly with the theoretic result $|W_0|=0.25-V_0$. }
 	\label{fig-sp}
 \end{figure*}

\section{Nonlinear modes in the $\PT$-symmetric Scarff-II potential}

The first potential to consider is the celebrated $\PT$-symmetric Scarff-II potential~\cite{real4}
\bee\label{ps}
 V(x)=V_0{\rm sech}^2x, \quad
 W(x)=W_0{\rm sech}x\tanh x,
\ene
with the real parameters $V_0<0$ and $W_0$ modulating the amplitudes of the reflectionless potential $V(x)$ and gain-and-loss distribution $W(x)$, respectively. For the case $W_0>0$, $W(x)$ represents the gain (loss) action in the domain of $x\geq 0$\, ($x\leq 0$), respectively, whereas $W_0<0$, $W(x)$ represents the gain (loss) action in the domain of $x\leq 0$\, ($x\geq 0$), respectively. Evidently, both $V(x)$ and $W(x)$ are bounded and vanish as $|x|\to \infty$. Moreover, the gain-and-loss distribution $W(x)$ always has a global balance in Eq.~(\ref{dnls}) since $\int^{+\infty}_{-\infty}W(x)dx=0$.

\subsection{Linear spectral problem}

In the absence of the derivative nonlinearity ($g=0$), Eq.~(\ref{ode}) becomes the following linear eigenvalue problem with the Scarff-II potential (\ref{ps})
\bee \label{lp}
 L\Phi(x)=\lambda\Phi(x),\quad L=-\partial_x^2+ V(x)+ iW(x),\,\,
\ene
with $\lambda$ and $\Phi(x)$ being the eigenvalue and localized eigenfunction, respectively. By virtue of the spectral method, we numerically find its symmetry-breaking line in $(V_0, W_0)$-space, which coincides well with the theoretical result that Eq.~(\ref{lp}) with Eq.~(\ref{ps}) enjoys entirely real spectra provided that $|W_0|\leq -V_0+1/4$~\cite{real4} (see Fig.~\ref{fig-sp}a). Therefore, for a fixed $W_0$ satisfying $|W_0|>1/4$, there always exists a threshold of the potential amplitude $V_0$, beyond which a phase transition occurs and the corresponding spectra become complex in the meantime (see Figs.~\ref{fig-sp}b, c).

 \begin{figure}[!t]
 	\begin{center}
 	\vspace{0.05in}
 	\hspace{-0.05in}{\scalebox{0.5}[0.5]{\includegraphics{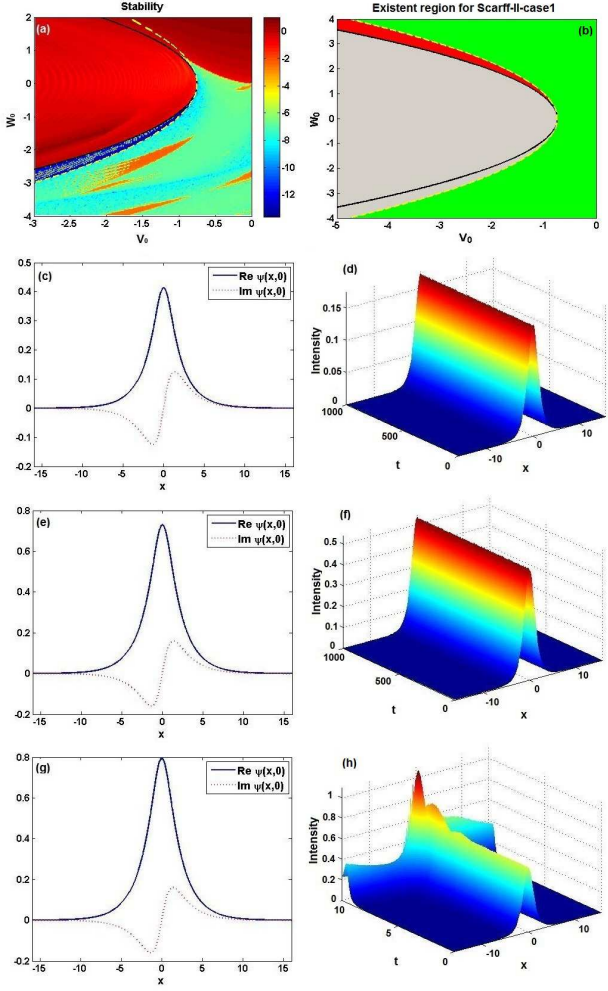}}}
 	\end{center}
 	\vspace{-0.15in} \caption{\small (color online).
   (a) Stable and unstable regions of nonlinear modes (\ref{sol-s}) [determined by the maximal absolute value of imaginary parts of the linearized eigenvalue $\delta$ in Eq.~(\ref{st}) in the $(V_0,W_0)$ space (common logarithmic scale), similarly hereinafter], where the yellow dashed parabola is $W_0^2+4V_0+3=0$, and black solid parabola is $4W_0^2+12V_0+9=0$. (b) The existence region (red and green) for Scarff-II-Case-1. Profile and evolution of nonlinear modes for (c, d) $V_0=-1,W_0=-1.1$ (unbroken linear $\PT$-symmetry), (e, f) $V_0=-1,W_0=-1.4$ (broken linear $\PT$-symmetry), (g, h) $V_0=-1,W_0=-1.5$ (broken linear $\PT$-symmetry).
 }
 	\label{stability-s-1}
 \end{figure}


However, more interestingly, even though the phase transition occurs in the linear spectral problem (i.e., Eq.~(\ref{lp}) has the complex spectra), nonlinear modes can still exist with entirely real eigenvalues, since the beam itself can have a strong influence on the amplitude of the potential through the derivative nonlinearity. Thus for the same parameter $W_0$, the new effective potential with stronger derivative nonlinearity may alter the linear $\PT$-symmetric threshold with the result that nonlinear eigenmodes can be found with real eigenvalues. But the broken $\PT$ symmetry cannot be nonlinearly restored at the lower power levels subject to the weaker derivative nonlinearity. Thus in what follows we turn to investigate nonlinear modes of Eq.~(\ref{dnls}) with $\PT$-symmetric Scarff-II potential (\ref{ps}) analytically and numerically.

 \begin{figure}[!t]
 	\begin{center}
 	\vspace{0.05in}
 	\hspace{-0.05in}{\scalebox{0.5}[0.5]{\includegraphics{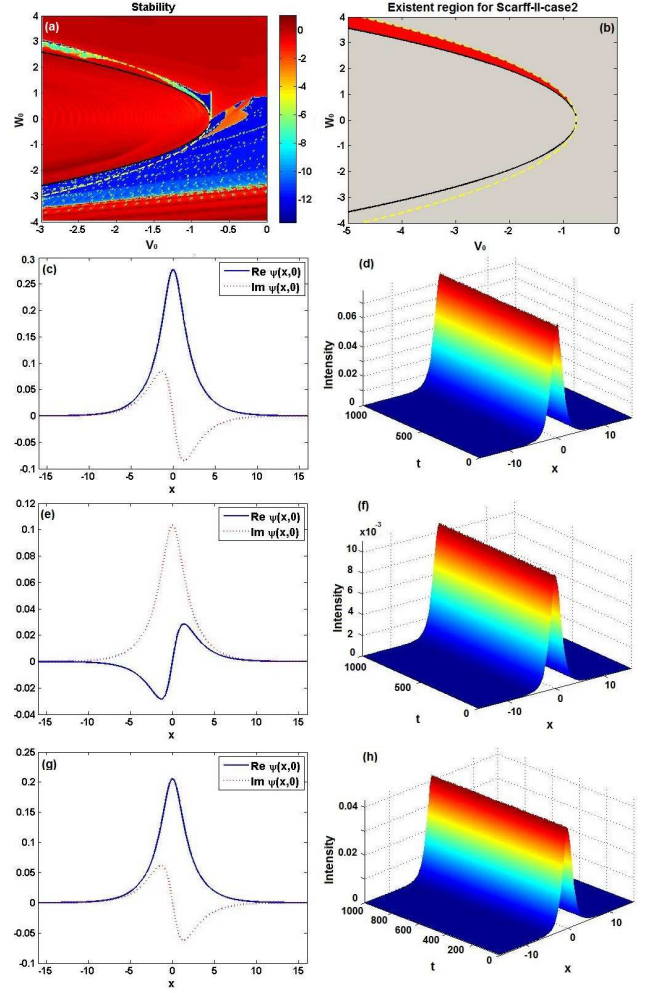}}}
 	\end{center}
 	\vspace{-0.15in} \caption{\small (color online).
   (a) Stable and unstable regions of nonlinear modes (\ref{sol-s}), where the yellow dashed parabola is $W_0^2+4V_0+3=0$, and the black solid parabola is $4W_0^2+12V_0+9=0$. (b) The existence region (only red) for Scarff-II-Case-2. Profile and evolution of nonlinear modes with unbroken linear $\PT$-symmetry for (c, d) $V_0=-0.9,W_0=0.74$ (exact solution), (e, f) $V_0=-0.9,W_0=0.78$ (inexact solution), (g, h) $V_0=-0.91,W_0=0.78$ (exact solution).}
 	\label{stability-s-2}
 \end{figure}

\subsection{Stability and dynamics of nonlinear modes}

Without loss of generality, we consider $g=1$. The exact bright solitons of Eq.~(\ref{ode}) with the Scarff-II potential (\ref{ps}) can be found in the form
\be\label{sol-s}
\phi(x)=\sqrt{\frac23\phi_0\,{\rm sech}x}\,\exp[i\varphi(x)],
\ee
where $\phi_0=W_0\pm\sqrt{4W_0^2+12V_0+9}>0$ (`$+$' denotes Scarff-II-Case-1 and `$-$' Scarff-II-Case-2, hereafter), the propagation constant is $\mu=0.25$, the nontrivial phase is $\varphi(x)=-\frac{(W_0+\phi_0)}2 {\rm tan}^{-1}[{\rm sinh}x]$ .

The existence conditions for the bright solitons (\ref{sol-s}) are
\bes\bee
 V_0>-\frac14(W_0^2+3) \,\,\,\, {\rm for} \,\,\,\, W_0<0 \qquad \vspace{0.05in} \\
 {\rm or} \qquad\qquad\qquad\qquad\qquad\qquad\qquad\qquad\qquad\qquad \nonumber\\
 V_0>-(\frac{W_0^2}{3}+\frac34) \,\,\,\, {\rm for} \,\,\,\, W_0>0 \qquad
 \ene
 \ees
  for the Scarff-II-Case-1 and
 \bee
  -(\frac{W_0^2}{3}+\frac34)\leq V_0<-\frac14(W_0^2+3) \,\,\,\, {\rm for} \,\,\,\,  W_0>0
 \ene
 for the Scarff-II-Case-2.
  Apparently, the nonlinear localized modes (\ref{sol-s}) are also $\PT$-symmetric. It is easy to see that for the same $\PT$-symmetric potential, the solutions (\ref{sol-s}) of the DNLS equation and ones of NLS equation (see Refs.~\cite{yanpre15,mu08}) have the distinct properties.

It is easy to see from Fig.~\ref{fig-sp}a that except for the only one tangent point $(V_0, W_0)=(-0.75, 0)$, the dashed parabola $V_0=-(W_0^2/3+0.75)$ is completely contained in the solid parabola $V_0=-0.25(W_0^2+3)$, which is tangent with the two linear $\PT$-symmetric breaking lines $\pm W_0=0.25-V_0$ with two tangent points being $(V_0, W_0)=(-1.75,\pm2)$. Thus the existence region of bright solitons (\ref{sol-s}) for Scaff-II-Case-1 contains both entire region of broken $\PT$-symmetric phase and partial region of unbroken $\PT$-symmetric phase (see Fig.~\ref{stability-s-1}b), whereas the existence region of  bright solitons (\ref{sol-s}) for Scaff-II-Case-2 is only located between the two parabolas in upper half plane, utterly located in the region of unbroken linear $\PT$-symmetric phase (see Fig.~\ref{stability-s-2}b). Moreover, we find that the strength $V_0$ and $W_0$ of the potential (\ref{ps}) can modulate not only amplitudes of bright solitons (\ref{sol-s}) but also the corresponding power $P=\int_{-\infty}^{+\infty}{|\psi(x,t)|^2}dx=2\pi\phi_0/3$, which is conserved.

In the following we investigate numerically the linear stability of bright solitons (\ref{sol-s}) for the Scarff-II-Case-1 and Scarff-II-Case-2 through the direct wave propagation of initially stationary modes (\ref{sol-s}) with some $2\%$ noise perturbation. Fig.~\ref{stability-s-1}a for Scarff-II-Case-1 and Fig.~\ref{stability-s-2}a for Scarff-II-Case-2 exhibit the stable (blue) and unstable (red) regions of nonlinear localized modes (\ref{sol-s}), respectively, which are determined by the maximum absolute value of imaginary parts of the linearized eigenvalue $\delta$ in Eq.~(\ref{st}) in $(V_0, W_0)$-space. For Scarff-II-Case-1 with $V_0=-1,W_0=-1.1$, belonging to the region of unbroken linear $\PT$-symmetric phase (see Fig.~\ref{fig-sp}a), the corresponding nonlinear localized mode is stable (see Figs.~\ref{stability-s-1}d). If we fix $V_0=-1$ and change $W_0=-1.4$ (it in fact holds for $W_0\in (-1.25, -1.4]$), in spite of belonging to the region of broken linear $\PT$-symmetric phase, the corresponding nonlinear localized mode can still keep stable (see Fig.~\ref{stability-s-1}g), that is, the derivative nonlinearity can excite the broken linear $\PT$-symmetric phase to the unbroken nonlinear $\PT$-symmetric phase. If we further increase $W_0$ a little bit to $W_0=-1.5$ (broken $\PT$-symmetric phase), the corresponding nonlinear mode begins to grow to become unstable (see Fig.~\ref{stability-s-1}h).

For the Scarff-II-Case-2, the bright solitons (\ref{sol-s}) only exist in the extremely narrow region between those two parabolas contained in the domain of unbroken $\PT$-symmetric phase (Fig.~\ref{stability-s-2}b). We find the stable nonlinear mode for $V_0=-0.9,\, W_0=-0.74$ (Fig.~\ref{stability-s-2}d). When we fix $V_0=-0.9$ and increase $W_0$ to $W_0=0.78$, in which
the solution becomes $\phi_{in}(x)=ia  \sqrt{{\rm sech}x}\, {\rm exp}[-ib {\rm tan}^{-1}({\rm sinh}(x))]$ with $a=0.1032471136, b=0.3820050252$. The stationary function $\phi_{in}(x)$ does not solve Eq.~(\ref{ode}) and its real (imaginary) part is an odd (even) function differing from the former cases, but we surprisedly find it can be stable through the direct evolution using the inexact solution $\phi_{in}(x)$ as an initial solution with some $2\%$ noise perturbation (see Fig.~\ref{stability-s-2}f).
When we fix $W_0=0.78$ and decrease $V_0$ a little bit to $V_0=-0.91$, in which the solution satisfies  Eq.~(\ref{dnls}) and the linear $\PT$-symmetric phase is unbroken, a stable nonlinear localized mode is found again (see Fig.~\ref{stability-s-2}h).

 \begin{figure}[!t]
 	\begin{center}
 	\vspace{0.05in}
 	\hspace{-0.05in}{\scalebox{0.4}[0.4]{\includegraphics{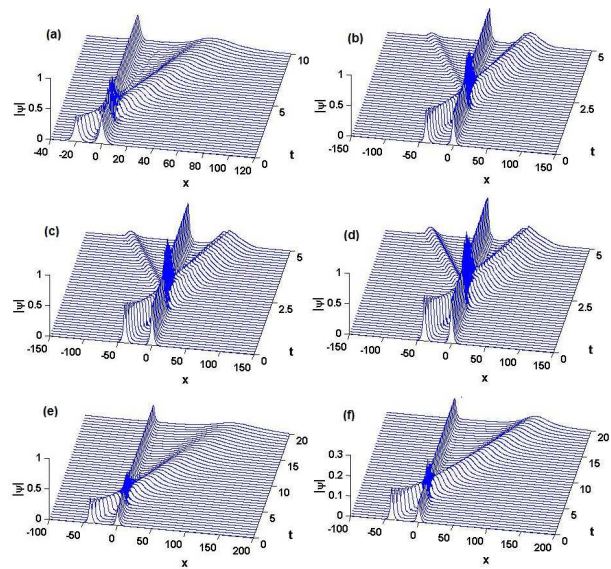}}}
 	\end{center}
 	\vspace{-0.15in} \caption{\small (color online).
    Interactions of two solitary waves in Eq.~(\ref{dnls}) with the Scarff-II potential (\ref{ps}). (a) The solution (\ref{sol-s}) for Scarff-II-Case-1 with the wave $\sqrt{\frac23 \phi_0 {\rm sech}(x+20)} e^{4i x}$ with $ V_0=-1, W_0=-1.1$. For $V_0=-1$, the solution (\ref{sol-s}) for Scarff-II-Case-1 with the wave $\sqrt{\frac23 \phi_0 {\rm sech}(x+40)} e^{10i x}$ with (b) $W_0=-1.2$, (c) $W_0=-1.3$, (d) $W_0=-1.4$. (e)  The solution (\ref{sol-s}) for Scarff-II-Case-2 with the wave $\sqrt{\frac32 \phi_0 {\rm sech}(x+40)} e^{4ix}$ with with $V_0=-0.9, W_0=0.74$. (f) The solution (\ref{sol-s}) for Scarff-II-Case-2 with the wave $\sqrt{\frac32 \phi_0 {\rm sech}(x+40)} e^{4ix}$ with with $V_0=-0.9, W_0=0.78$.}
 	\label{collision-s-12}
 \end{figure}

 \begin{figure}[!t]
 	\begin{center}
 	\vspace{0.05in}
 	\hspace{-0.05in}{\scalebox{0.4}[0.4]{\includegraphics{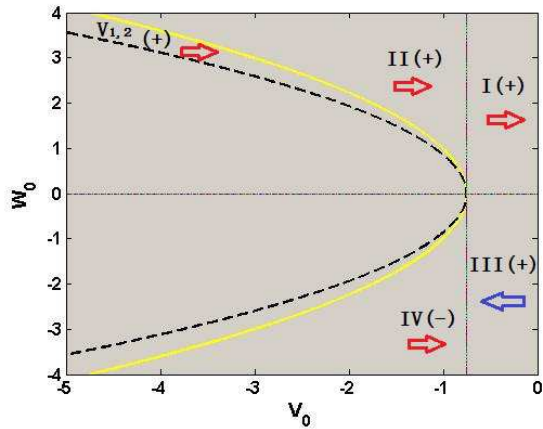}}}
 	\end{center}
 	\vspace{-0.15in} \caption{\small (color online).
   Signs and directions of the transverse power flow $S(x)$ with regard to nonlinear modes (\ref{sol-s}). The solid parabola $W_0^2+4V_0+3=0$, dashed parabola is $4W_0^2+12V_0+9=0$, the horizontal line $W_0=0$ and vertical line $V_0=-3/4$ divide the existent region of exact soliton solutions (\ref{sol-s}) into six small domains (I,\,II,\,III,\,IV,\,V$_{1,2}$ for Scarff-II-Case-1, and only V$_{1,2}$ for Scarff-II-Case-2, where '$+$' ('$-$') denotes the positive (negative) sign of $S(x)$, and the red right (blue left) arrow denotes the direction of power flow from loss (gain) to gain (loss).}
 	\label{pf-s}
 \end{figure}

We now investigate the interaction between two solitary waves in the $\PT$-symmetric Scarff-II potential. For the Scarff-II-Case-1 and $V_0=-1,\, W_0=-1.1$, we consider the initial condition $\psi(x,0)=\phi(x)+\sqrt{\frac23 \phi_0 {\rm sech}(x+20)} e^{4i x}$ with $\phi(x)$ determined by Eq.~(\ref{sol-s}), as a result, the semi-elastic interaction is generated in which exact nonlinear mode does not change its shape whereas the exotic incident wave becomes damped before and after interaction (see Fig.~\ref{collision-s-12}a). When $W_0$ becomes a little bit to $W_0=-1.4$, we consider the initial condition $\psi(x,0)=\phi(x)+\sqrt{\frac23 \phi_0 {\rm sech}(x+40)} e^{10i x}$ with $\phi(x)$ determined by Eq.~(\ref{sol-s}), then a novel phenomenon occurs in collision that there exists a reflected wave when exotic incident wave interacts with the exact soliton (\ref{sol-s}) (see Fig.~\ref{collision-s-12}d). Through repeated numerical tests, we find the reflected wave is probably related to the simultaneously increasing amplitude of the exact soliton and exotic incident wave. As $W_0$ decreases from $-1.1$ to $-1.4$, it is easy to verify that the amplitude (determined by $\phi_0$) of the exact soliton or exotic incident wave increases and in the meantime the reflected wave begins to occur and then becomes larger and larger (see Figs.~\ref{collision-s-12}(a, b, c, d)). However, the exact nonlinear mode still does not change its shape before and after interaction. Similarly, for the Scarff-II-Case-2, we successively consider the initial condition $\psi(x,0)=\phi(x)+\sqrt{\frac32 \phi_0 {\rm sech}(x+40)} e^{4ix}$ for $V_0=-0.9,W_0=0.74$ and $V_0=-0.9,W_0=0.78$ with $\phi(x)$ determined by Eq.~(\ref{sol-s}), the similar semi-elastic interactions to Fig.~\ref{collision-s-12}a are generated  (see Figs.~\ref{collision-s-12}(e, f)).

In order to better understand the properties of the nonlinear localized modes (\ref{sol-s}), we check its corresponding transverse power flow (Poynting vector), which derives from the nontrivial phase structure of the nonlinear localized modes and is given by $S(x)=\frac{i}2(\psi\psi_x^{\ast}-\psi^{\ast}\psi_x)=-\frac13\phi_0 (W_0+\phi_0){\rm sech}^2x$ with $\phi_0>0$. Signs and directions of the transverse power flow $S(x)$ are discussed and summarized in detail in Fig.~\ref{pf-s}.

\subsection{Excitations of nonlinear modes}

 \begin{figure}[!t]
 	\begin{center}
 	\vspace{0.05in}
 	\hspace{-0.05in}{\scalebox{0.4}[0.4]{\includegraphics{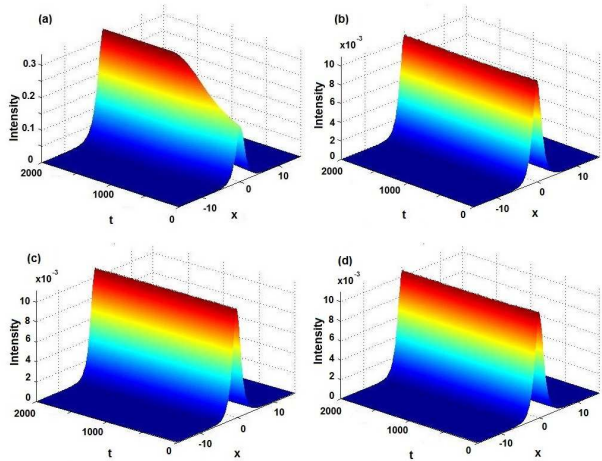}}}
 	\end{center}
 	\vspace{-0.15in} \caption{\small (color online). Excitation of stable nonlinear localized states [cf. Eq.~(\ref{tdnls})]. (a) $V_0= -1, W_{01}= -1.1, W_{02}= -1.4$, from stable nonlinear state belonging  to unbroken linear $\PT$-symmetry to stable nonlinear state belonging to broken linear $\PT$-symmetry for Scarff-II-Case-1; (b) $V_0= -0.9, W_{01}= 0.78, W_{02}= 0.74$, (c) $V_{01}= -0.9, V_{02}= -0.91,, W_0= 0.78$, (d) $V_{01}= -0.9, V_{02}= -0.91, W_{01}= 0.78, W_{02}= 0.74$, from stable inexact nonlinear state to stable exact nonlinear state with unbroken linear $\PT$-symmetry for Scarff-II-Case-2. }
 	\label{excited-s-12}
 \end{figure}

Finally, we discuss the excitation of nonlinear localized modes by means of changing the potential amplitudes as the functions of time, $V_0\rightarrow V_0(t)$ or $W_0\rightarrow W_0(t)$ [cf. Ref.~\cite{yanpre15}]. It means that we focus on the simultaneous adiabatic switching on the Scarff-II potential, governed by
\be\label{tdnls}
i\psi_t+\psi_{xx}-[V(x,t)+iW(x,t)]\psi+ig(|\psi|^2\psi)_x=0,
\ee
where $V(x,t), W(x,t)$ are given by Eq.~(\ref{ps}) with $V_0\rightarrow V_0(t)$ and $W_0\rightarrow W_0(t)$, and $V_0(t)$, $W_0(t)$ are both choose as the following form
\be\label{excite-s}\epsilon(t)=
\begin{cases}
(\epsilon_2-\epsilon_1) \sin(\pi t/2000)+\epsilon_1, & \text{$0\leq t<1000$},\\
\epsilon_2, & \text{$t\geq 1000$}
\end{cases}
\ee
where $\epsilon_{1,2}$ are real constants. It is easy to verify that nonlinear localized modes (\ref{sol-s}) with $V_0\rightarrow V_0(t)$ or $W_0\rightarrow W_0(t)$ do not satisfy Eq.~(\ref{tdnls}) any more, whereas the modes (\ref{sol-s}) do satisfy Eq.~(\ref{tdnls}) for both the initial state $t=0$ and excited states $t\geq 1000$.

For the Scarff-II-Case-1, Fig.~\ref{excited-s-12}a exhibits the wave propagation of the nonlinear modes $\psi(x,t)$ of Eq.~(\ref{tdnls}) via the initial condition given by Eq.~(\ref{sol-s}) with $W_0\rightarrow W_0(t)$ given by Eq.~(\ref{excite-s}), which excite an initially stable nonlinear localized mode given by Eq.~(\ref{sol-s}) for $(V_0, W_{01})=(-1, -1.1)$ with the unbroken linear $\PT$-symmetric phase to another stable nonlinear localized mode given by Eq.~(\ref{sol-s}) for $(V_0, W_{02})=(-1, -1.4)$, though with broken linear $\PT$-symmetric phase. It also indicates fully that bright solitons (\ref{sol-s}) have extremely strong capacity of resisting disturbance.

For the Scarff-II-Case-2, we successively perform three types of excitations by changing  potential amplitudes $V_0\rightarrow V_0(t)$ or $W_0\rightarrow W_0(t)$ singly or simultaneously. Similarly, Fig.~\ref{excited-s-12}b displays the wave propagation of nonlinear modes $\psi(x,t)$ of Eq.~(\ref{tdnls}) using the initial condition given by Eq.~(\ref{sol-s}) with $W_0\rightarrow W_0(t)$ given by Eq.~(\ref{excite-s}), which excites a stable and inexact nonlinear localized mode given by Eq.~(\ref{sol-s}) for $(V_0, W_{01})=(-0.9, 0.78)$ to another stable and exact nonlinear localized mode given by Eq.~(\ref{sol-s}) for $(V_0, W_{02})=(-0.9, 0.74)$, both of which belong to unbroken linear $\PT$-symmetric phase. The property is fairly significant to find  numerically or experimentally the stable nonlinear mode from an inexact mode due to the stability of excitation. Of course, we can also achieve the same purpose only by tuning $V_0$ appropriately. Fig.~\ref{excited-s-12}c displays the wave propagation of the nonlinear modes $\psi(x,t)$ of Eq.~(\ref{tdnls}) via the initial condition given by Eq.~(\ref{sol-s}) with $V_0\rightarrow V_0(t)$ given by Eq.~(\ref{excite-s}), which excites a stable and inexact nonlinear localized mode given by Eq.~(\ref{sol-s}) for $(V_{01}, W_0)=(-0.9, 0.78)$ to another stable and exact nonlinear localized mode given by Eq.~(\ref{sol-s}) for $(V_{02}, W_0)=(-0.91, 0.78)$, both of which belong to unbroken linear $\PT$-symmetric phase.  Fig.~\ref{excited-s-12}d shows the wave propagation of nonlinear modes $\psi(x,t)$ of Eq.~(\ref{tdnls}) via the initial condition given by Eq.~(\ref{sol-s}) with $V_0\rightarrow V_0(t), W_0\rightarrow W_0(t)$ given by Eq.~(\ref{excite-s}), which excites a stable and inexact nonlinear localized mode given by Eq.~(\ref{sol-s}) for $(V_{01}, W_{01})=(-0.9, 0.78)$ to another stable and exact nonlinear localized mode given by Eq.~(\ref{sol-s}) for $(V_{02}, W_{02})=(-0.91, 0.74)$, both of which also belong to unbroken linear $\PT$-symmetric phase.

\section{Nonlinear modes in the $\PT$-symmetric harmonic-Hermite-Gaussian potential}

Next we consider another physically significant potential, that is, the harmonic potential and gain-and-loss distribution of Hermite-Gaussian type
\bee\label{phv}
V(x)=\omega^2 x^2, \qquad\qquad\qquad\qquad \qquad\quad\\ \label{phw}
W_n(x)=\sigma H_n(\sqrt{\omega}x)[\omega x H_n(\sqrt{\omega}x) \qquad\quad \notag \\
 \qquad\quad -2n\sqrt{\omega}H_{n-1}(\sqrt{\omega}x)]e^{-\omega x^2},
\ene
where the frequency $\omega>0$ and real constant $\sigma>0$ can adjust amplitudes of the harmonic potential $V(x)$ and gain-and-loss distribution $W_n(x)$, and $H_n(x)=(-1)^n e^{x^2}(d^n e^{-x^2})/(dx^n)$ represents the Hermite polynomial with $n$ being a non-negative integer and $H_n(x)\equiv0$ as $n<0$. It is easy to verify that these complex potentials $V(x)+iW_n(x)$ are all $\PT$-symmetric for any non-negative integer $n$, which differ from other ones~\cite{yanpre15}. Without loss of generality, in what follows we mainly focus on the $\PT$-symmetric potentials (\ref{phv}) and (\ref{phw}) for $n=0,1,2$.

 \begin{figure}[!t]
 	\begin{center}
 	\vspace{0.05in}
 	\hspace{-0.05in}{\scalebox{0.5}[0.5]{\includegraphics{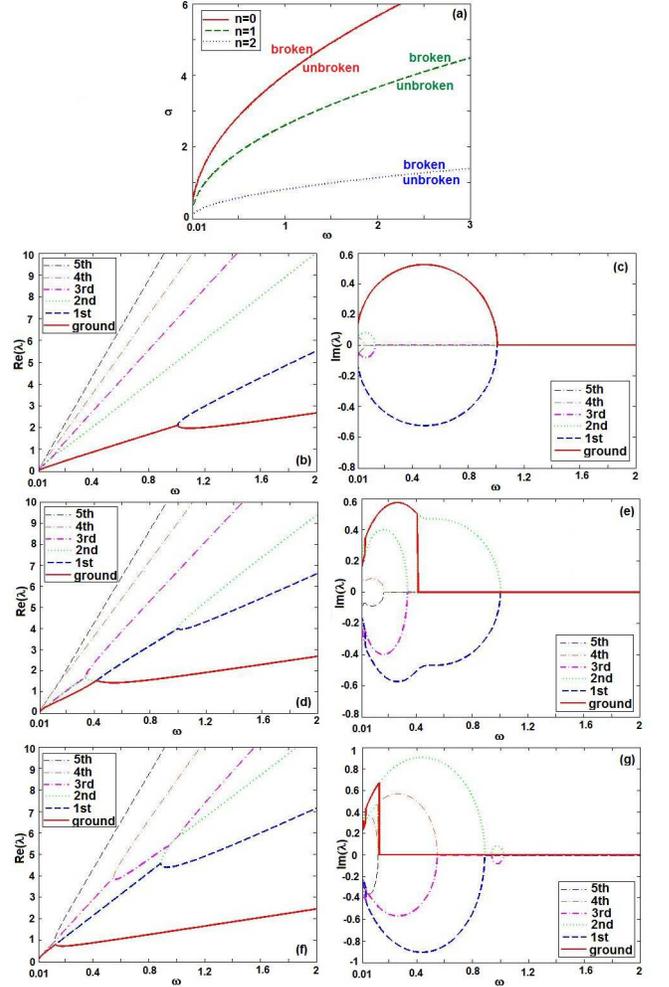}}}
 	\end{center}
 	\vspace{-0.15in} \caption{\small (color online).
   (a) The unbroken (broken) $\PT$-symmetric phase is in the domain below (above) the phase-breaking curves ($n=0$: red solid, $n=1$: green dashed, $n=2$: blue dotted) for the linear operator $L$ in Eq.~(\ref{lp}) with $\PT$-symmetric harmonic potentials (\ref{phv}) and gain-and-loss distributions (\ref{phw}). (b, d, f) Real and (c, e,g) imaginary parts of the eigenvalues $\lambda$ of the linear problem (\ref{lp}) with $\PT$-symmetric potential (\ref{phv}) and (\ref{phw}) as a function of $\omega$, at ($\sigma=4$,\, $n=0$), ($\sigma=2.59$,\, $n=1$), and ($\sigma=0.81$,\,  $n=2$), respectively. The three phase transition thresholds are all approximately $\omega=1$, in accord with the phase-breaking curves in (a).}
 	\label{spectra-g-012}
 \end{figure}

 \begin{figure*}[!t]
 	\begin{center}
 	\vspace{0.05in}
 	\hspace{-0.05in}{\scalebox{0.8}[0.8]{\includegraphics{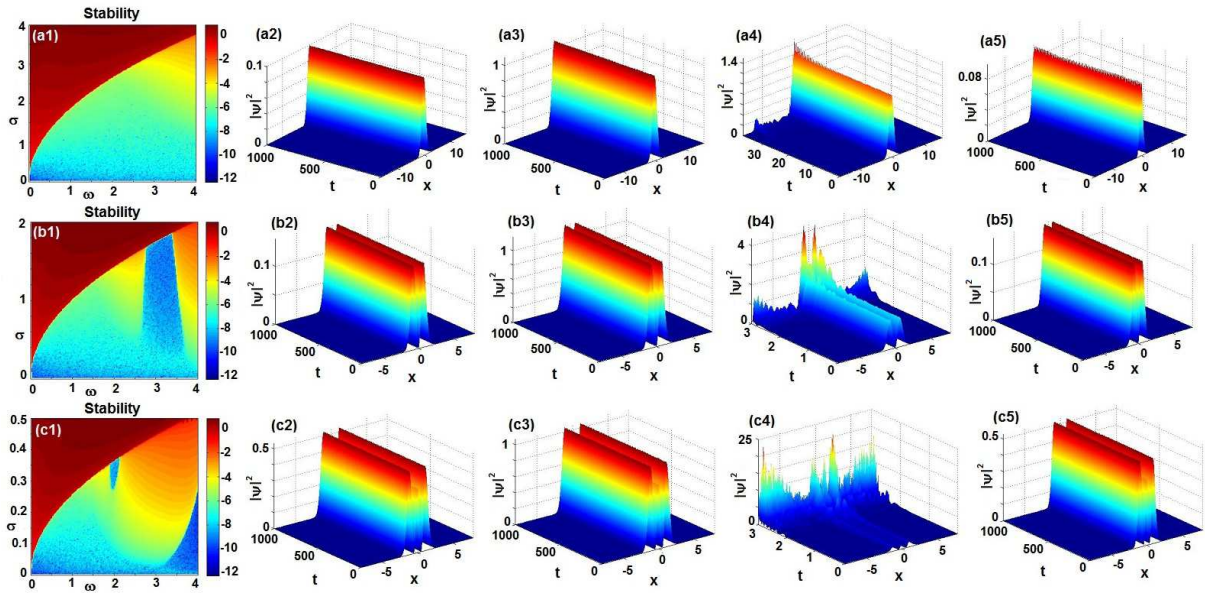}}}
 	\end{center}
 	\vspace{-0.15in} \caption{\small (color online). Linear stability of nonlinear modes (\ref{sol-h}) for (a1) $n=0$, (b1) $n=1$, and (c1) $n=2$. Evolutions of nonlinear modes (\ref{sol-h}) for one-hump ($n=0$) with unbroken linear $\PT$-symmetry [(a2) $\omega=1,\, \sigma=0.1$ (stable), (a3) $\omega=1,\, \sigma=1.1$ (stable), (a4) $\omega=1,\, \sigma=1.2$ (unstable), (a5) $\omega=2,\, \sigma=0.1$ (periodically varying)], two-hump ($n=1$) with unbroken linear $\PT$-symmetry [(b2) $\omega=2,\, \sigma=0.1$ (stable), (b3) $\omega=2,\, \sigma=0.8$ (stable), (b4) $\omega=2,\, \sigma=1$ (unstable), (b5) $\omega=3,\, \sigma=0.1$ (stable)], three-hump ($n=2$) with unbroken linear $\PT$-symmetry [(c2) $\omega=2,\, \sigma=0.1$ (stable), (c3) $\omega=2,\, \sigma=0.2$ (stable), (c4) $\omega=2,\, \sigma=0.3$ (unstable), (c5) $\omega=3,\, \sigma=0.1$ (stable)]. }
 	\label{sta-evo-g-012}
 \end{figure*}

\subsection{Linear spectral problem}

Here we investigate the linear operator $L$ in Eq.~(\ref{lp}) with $\PT$-symmetric potentials composed of $V(x)$ (\ref{phv}) and $W_{0,1,2}(x)$ (\ref{phw}), which are explicitly given by ($n=0, 1,2)$
\bes\label{W0}
\bee
\text{$W_0(x)=\sigma \omega x e^{-\omega x^2}$},  \label{W1} \\
\text{$W_1(x)=4\sigma \omega x(\omega x^2-1) e^{-\omega x^2}$},  \label{W2} \\
\text{$W_2(x)=4\sigma \omega x(4\omega^2 x^4-12\omega x^2+5) e^{-\omega x^2}$},
\ene
\ees

For $n=0,1,2$, the regions of unbroken and broken linear $\PT$-symmetric phase on $(\omega, \sigma)$-space are all numerically exhibited in Fig.~\ref{spectra-g-012}. It can be obviously observed that the ranges of unbroken linear $\PT$-symmetric phase gradually shrink with $n$ increasing, which is mainly because the higher amplitude of gain-and-loss distribution $W_n(x)$ can possibly lead to the broken linear $\PT$-symmetric phase as $n$ increases. For some fixed $\sigma$, we also illustrate numerically the collisions of the first six lowest discrete energy levels as the frequency $\omega$ decreases (see Figs.~\ref{spectra-g-012}(b-g)). Notice that only the first two lowest energy levels interact with each other for $n=0$ whereas the situations become more and more intricate with $n$ growing.

\subsection{Nonlinear modes and stability}

For the above-mentioned $\PT$-symmetric potential $V(x)+iW_n(x)$ with Eqs.~(\ref{phv}) and (\ref{phw}), we find a series of multi-hump bright solitons of Eq.~(\ref{ode})
\be\label{sol-h}
\phi_n(x)=\sqrt{\sigma} H_n(\sqrt{\omega} x) e^{-\omega x^2/2} e^{i\varphi(x)},\quad \sigma>0,
\ee
where the chemical potential $\mu=-\omega (2n+1)$, and the phase function $\varphi_n(x)=-\sigma \int_0^x{H_n^2(\sqrt{\omega}s)e^{-\omega s^2}} ds$.

For the cases $n=0,1,2$, we first give the regions of linear stability [cf. Eq.~(\ref{st})] of nonlinear localized modes (\ref{sol-h}) in the $(\omega, \sigma)$ space (see Figs.~\ref{sta-evo-g-012}(a1, b1, c1)). It is more than evident that the stable regions of linear stability have the similar narrowing behaviors to the corresponding unbroken $\PT$-symmetric phase above on account of the rising strength of the gain-and-loss distribution $W_n(x)$ as $n$ increases. Moreover, the stable regions of linear stability are entirely included in the regions of the corresponding unbroken $\PT$-symmetric phase, which indicates the derivative nonlinear term makes a negative influence on the corresponding linear $\PT$-symmetric phase.

We now study numerically the linear stability of bright solitons (\ref{sol-h}) for cases $n=0,1,2$ through the direct wave propagation of initially stationary mode (\ref{sol-h}) for several specific amplitude parameters $(\omega, \sigma)$ with some $2\%$ noise perturbation. For $n=0$ and the fixed $\omega=1$, we modulate $\sigma$ from a very small positive number (e.g.,
$\sigma=0.001$) to $\sigma=1.1$ to perform the direct wave evolution of  one-hump nonlinear modes (\ref{sol-h}) such that we obtain the stable one-hump solitons (see Figs.~\ref{sta-evo-g-012}(a2, a3)). Whereas we further increase $\sigma$ to $\sigma=1.2$, the one-hump nonlinear mode (\ref{sol-h}) begins to become extremely unstable (see Fig.~\ref{sta-evo-g-012}(a4)). The main reason is that the gain-and-loss distribution $W_n(x)$ has a stronger effect on the stability of modes as $\sigma$ increases. For the fixed $\sigma=0.1$, we only change $\omega$ from $1$ to $2$ continuously such that we also find a series of stable one-hump solitons, although there exist some small periodic variation as $\omega$ approaches to $2$ (see Fig.~\ref{sta-evo-g-012}(a5)).

 For $n=1$, we can also find a family of stable two-hump solitons (\ref{sol-h}) for a fixed $\omega=2$ and $\sigma=0.1\rightarrow 0.8$ (see Figs.~\ref{sta-evo-g-012}(b2, b3)). Whereas we further increase $\sigma$ from $0.8$ to $1$, the two-hump nonlinear mode (\ref{sol-h}) begins to become extremely unstable (see Fig.~\ref{sta-evo-g-012}(b4)). For a fixed $\sigma=0.1$, we change $\omega$ from $2$ to $3$ continuously such that we can also find a series of stable two-hump solitons (Fig.~\ref{sta-evo-g-012}(b5)).  For $n=2$, we also have the similar results for three-hump solitons (see Figs.~\ref{sta-evo-g-012}(c2-c5))

 \begin{figure}[!t]
 	\begin{center}
 	\vspace{0.05in}
 	\hspace{-0.05in}{\scalebox{0.4}[0.4]{\includegraphics{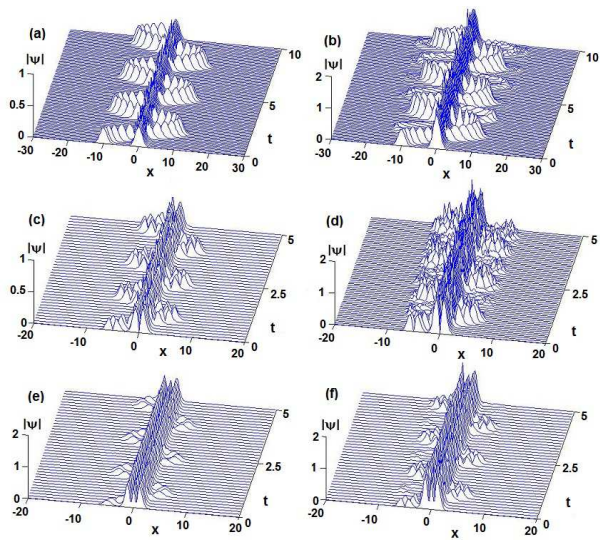}}}
 	\end{center}
 	\vspace{-0.15in} \caption{\small (color online). Interactions of two solitary waves in Eq.~(\ref{dnls}) with the potential (\ref{phv}) and (\ref{phw}). (a) The solution (\ref{sol-h}) for $n=0$ with the wave $0.8 \sqrt{\sigma} e^{-\omega (x+10)^2/2} e^{i \varphi(x)}$ with $\omega=1, \sigma=0.1$, (b) the solution (\ref{sol-h}) for $n=0$ with the wave $0.5 \sqrt{\sigma} e^{-\omega (x+10)^2/2} e^{i \varphi(x)}$ with $\omega=1, \sigma=1.1$, (c)  the solution (\ref{sol-h}) for $n=1$ with the wave $\sqrt{\sigma\omega} (x+5) e^{-\omega (x+5)^2/2} e^{i \varphi(x)}$ with $ \omega=2, \sigma=0.1$, (d) the solution (\ref{sol-h}) for $n=1$ with the wave $\sqrt{\sigma\omega} (x+5) e^{-\omega (x+5)^2/2} e^{i \varphi(x)}$ with $\omega=2, \sigma=0.8$, (e) the solution (\ref{sol-h}) for $n=2$ with the wave $0.5 \sqrt{\sigma} e^{-\omega (x+5)^2/2} e^{i \varphi(x)}$ with $ \omega=2, \sigma=0.1$, (f) the solution (\ref{sol-h}) for $n=2$ with the wave $\sqrt{\sigma\omega} (x+5) e^{-\omega (x+5)^2/2} e^{i \varphi(x)}$ with $\omega=2, \sigma=0.2$. }
 	\label{collision-h-012}
 \end{figure}

 \begin{figure}[!t]
 	\begin{center}
 	\vspace{0.05in}
 	\hspace{-0.05in}{\scalebox{0.42}[0.42]{\includegraphics{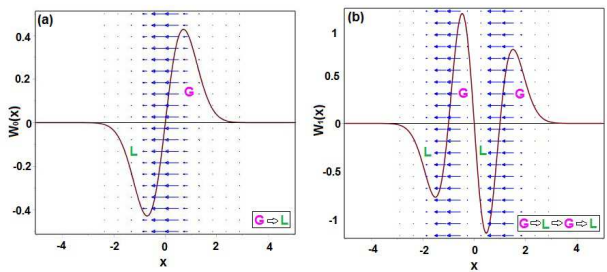}}}
 	\end{center}
 	\vspace{-0.15in} \caption{\small (color online). (a) The direction of the transverse power flow for $n=0$ is from the gain to loss. (b) The direction of the transverse power flow for $n=1$ is firstly from the gain to loss, then from the loss to gain, finally from the gain to loss again (from left to right). Here, $W_0(x)$ only has one root $x=0$ and $W_1(x)$ has three roots $x=0,\pm1$. The left arrows and its lengths denote respectively negative directions and strengths of the transverse power flow $S_{0,1}(x)$ with regard to nonlinear modes (\ref{sol-h}). 'G' ('L') denote the gain (loss) distribution of $W_{0,1}(x)$. Other parameters are $\omega=\sigma=1$. }
 	\label{pf-h}
 \end{figure}

 Next we investigate the interactions of bright solitons (\ref{sol-h}) in the $\PT$-symmetric potential $V(x)+iW_n(x)$ with (\ref{phv}) and (\ref{phw}). For $n=0$ and $\omega=1, \sigma=0.1$, we consider the initial condition $\psi(x,0)=\phi(x)+0.8 \sqrt{\sigma} e^{-\omega (x+10)^2/2} e^{i \varphi(x)}$ with $\phi(x)$ determined by Eq.~(\ref{sol-h}), as a result the elastic interaction is generated in which neither the exact one-hump nonlinear mode nor exotic periodic incident wave change their shapes before and after interaction (see Fig.~\ref{collision-h-012}a). When $\sigma$ becomes large to $\sigma=1.1$, we consider the initial condition $\psi(x,0)=\phi(x)+0.5 \sqrt{\sigma} e^{-\omega (x+10)^2/2} e^{i \varphi(x)}$ with $\phi(x)$ determined by Eq.~(\ref{sol-h}), then a novel phenomenon occurs in collision that there exists a weak reflected wave when exotic incident wave interacts with the exact one-hump soliton (\ref{sol-h}) (see Fig.~\ref{collision-h-012}b), which is probably related to the increasing amplitude or strength of the exact one-hump soliton (\ref{sol-h}). However, the exact one-hump nonlinear mode still doesn't change its shape before and after interaction. Similarly, for $n=1$, we successively consider the initial condition $\psi(x,0)=\phi(x)+\sqrt{\sigma\omega} (x+5) e^{-\omega (x+5)^2/2} e^{i \varphi(x)}$ for $\omega=2, \sigma=0.1$ and $\omega=2, \sigma=0.8$ with $\phi(x)$ determined by Eq.~(\ref{sol-h}), the similar elastic interactions between the exact two-hump nonlinear modes and exotic periodic incident waves to Fig.~\ref{collision-h-012}a are generated (see Figs.~\ref{collision-h-012}c, d). For $n=2$, the similar elastic interactions between the exact three-hump nonlinear modes and exotic periodic incident waves are generated (see Figs.~\ref{collision-h-012}e, f).

 In order to further understand the properties of the stationary nonlinear localized modes (\ref{sol-h}), we also check its corresponding transverse power flow (Poynting vector) $S_n(x)=-\sigma^2 H_n^4(\sqrt{\omega}x) e^{-2\omega x^2}$ with $\omega>0$ and $\sigma>0$. Notice that the signs of the transverse power flow $S_n(x)$ always keep negative definite for any $n$. For $n=0$, the power always flows in one direction, i.e.,
from the gain toward the loss domain (see Fig.~\ref{pf-h}a). However, the directions of the power flow for $n=1$ are so complicated, that is (from the negative infinite to positive infinite), firstly from the gain to loss, then from the loss to gain, finally from the gain to loss domain again (see Fig.~\ref{pf-h}b, the total direction is from the gain to loss). Similar more complicated results hold for $n=2$ and specifically not repeat them.

\subsection{Excitations of nonlinear modes}

Finally, we investigate the excitation of stable nonlinear modes by means of changing the potential amplitudes $\omega\rightarrow \omega(t)$ or $\sigma\rightarrow \sigma(t)$, which means that we exert simultaneous adiabatic switching on the harmonic potential (\ref{phv}) and gain-and-loss distribution (\ref{phw}), modelled by Eq.~(\ref{tdnls}), where $V(x,t), W(x,t)$ are given by Eq.~(\ref{phv}) and Eq.~(\ref{phw}) with $\omega\rightarrow \omega(t)$ and $\sigma\rightarrow \sigma(t)$. We assume that $\omega(t)$, $\sigma(t)$ are all taken as the same form as Eq.~(\ref{excite-s}) (i.e., $\epsilon(t)$ can be replaced with $\omega(t)$ or $\sigma(t)$). We have the similar results that the nonlinear localized modes (\ref{sol-h}) with $\omega\rightarrow \omega(t)$ or $\sigma\rightarrow \sigma(t)$ do not satisfy Eq.~(\ref{tdnls}) any more, whereas the modes (\ref{sol-h}) indeed satisfy Eq.~(\ref{tdnls}) for the initial state $t=0$ and excited states $t\geq 1000$.

\begin{figure}[!t]
 	\begin{center}
 	\vspace{0.05in}
 	\hspace{-0.05in}{\scalebox{0.42}[0.45]{\includegraphics{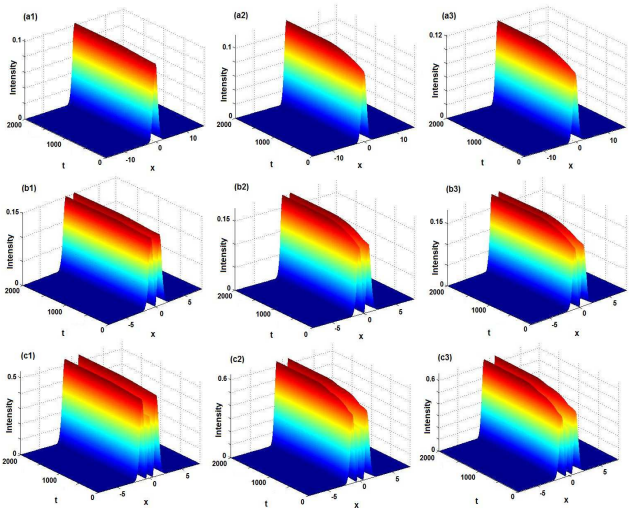}}}
 	\end{center}
 	\vspace{-0.15in} \caption{\small (color online). Excitation of stable nonlinear localized states [cf. Eq.~(\ref{tdnls})] for $n=0$ [(a1) $\omega= 1, \sigma_1= 0.1, \sigma_2= 1.1$, (a2) $\omega_1= 1, \omega_2= 2,\sigma= 0.1$, (a3) $\omega_1= 1, \omega_2= 2,\sigma_1= 0.1, \sigma_2= 1.1$]; $n=1$ [(b1) $\omega= 2, \sigma_1= 0.1, \sigma_2= 0.8$, (b2) $\omega_1= 2, \omega_2= 3,\sigma= 0.1$, (b3) $\omega_1= 2, \omega_2= 3,\sigma_1= 0.1, \sigma_2= 0.8$]; $n=2$ [(c1) $\omega= 2, \sigma_1= 0.1, \sigma_2= 0.2$, (c2) $\omega_1= 2, \omega_2= 3,\sigma= 0.1$, (c3) $\omega_1= 2, \omega_2= 3,\sigma_1= 0.1, \sigma_2= 0.2$].}
 \label{excited-h-012}
 \end{figure}

For $n=0,1,2$, we numerically perform three distinct types of excitations by changing the potential amplitudes $\omega\rightarrow \omega(t)$, $\sigma\rightarrow \sigma(t)$, or both, respectively. For $n=0$, Fig.~\ref{excited-h-012}(a1) exhibits the wave evolution of nonlinear modes $\psi(x,t)$ of Eq.~(\ref{tdnls}) via the initial condition given by Eq.~(\ref{sol-h}) with $\sigma\rightarrow \sigma(t)$ given by Eq.~(\ref{excite-s}), which excites a stable one-hump nonlinear localized mode given by Eq.~(\ref{sol-h}) for $(\omega, \sigma_1)=(1, 0.1)$ to another stable one-hump nonlinear localized mode given by Eq.~(\ref{sol-h}) for $(\omega, \sigma_2)=(1, 1.1)$, which both belong to unbroken linear $\PT$-symmetric phase (all mentioned points $(\omega, \sigma)$ enjoy the same property of unbroken linear $\PT$-symmetric phase hereafter). Fig.~\ref{excited-h-012}(a2) displays the wave propagation of the nonlinear modes $\psi(x,t)$ of Eq.~(\ref{tdnls}) via the initial condition given by Eq.~(\ref{sol-h}) with $\omega\rightarrow \omega(t)$ singly given by Eq.~(\ref{excite-s}), which excite a initially stable one-hump nonlinear localized mode given by Eq.~(\ref{sol-h}) for $(\omega_1, \sigma)=(1, 0.1)$ to another stable one-hump nonlinear localized mode given by Eq.~(\ref{sol-h}) for $(\omega_2, \sigma)=(2, 0.1)$. Fig.~\ref{excited-h-012}(a3) displays the wave propagation of the nonlinear modes $\psi(x,t)$ of Eq.~(\ref{tdnls}) via the initial condition given by Eq.~(\ref{sol-h}) with $\omega\rightarrow \omega(t)$ and $\sigma\rightarrow \sigma(t)$ simultaneously given by Eq.~(\ref{excite-s}), which excite a stable one-hump nonlinear localized mode given by Eq.~(\ref{sol-h}) for $(\omega_1, \sigma_1)=(1, 0.1)$ to another stable one-hump nonlinear localized mode given by Eq.~(\ref{sol-h}) for $(\omega_2, \sigma_2)=(2, 1.1)$. Similarly for $n=1$, we excite a stable two-hump nonlinear localized mode given by Eq.~(\ref{sol-h}) for $(\omega, \sigma)=(2, 0.1)$ to another stable two-hump nonlinear localized mode given by Eq.~(\ref{sol-h}) for $(\omega, \sigma)=(2, 0.8)$, $(\omega, \sigma)=(3, 0.1)$, and $(\omega, \sigma)=(3, 0.8)$, respectively (see Figs.~\ref{excited-h-012}(b1, b2, b3)). For $n=2$, we also excite a stable three-hump nonlinear localized mode given by Eq.~(\ref{sol-h}) for $(\omega, \sigma)=(2, 0.1)$ to another stable three-hump nonlinear localized mode given by Eq.~(\ref{sol-h}) for $(\omega, \sigma)=(2, 0.2)$, $(\omega, \sigma)=(3, 0.1)$, and $(\omega, \sigma)=(3, 0.2)$, respectively (see Figs.~\ref{excited-h-012}(c1, c2, c3)). These stable excitations also shows that bright solitons (\ref{sol-h}) have extremely strong capacity of resisting disturbance.

\section{Conclusions and discussions}

In conclusion, some stable bright solitons have been investigated in the derivative nonlinear Schr\"odinger equation with $\PT$-symmetric Scarff-II and hamonic-Hermite-Gaussian potentials. Firstly, the linear $\PT$-symmetric breaking curves are numerically exhibited. Secondly, in the presence of derivative nonlinearity, such $\PT$-symmetric solitons are shown to be stable through the linear stability analysis and direct wave propagation with some noise perturbation. Moreover, the semi-elastic interactions between exact bright solitons and exotic incident waves are illustrated and the transverse power flows are also checked in detail. Finally, the soliton excitations are also studied including from a stable nonlinear mode with unbroken linear $\PT$-symmetric phase to another stable nonlinear mode with broken linear $\PT$-symmetric phase and from a stable and inexact nonlinear mode to another stable and exact nonlinear mode. In fact, we may change the nonlinear coefficient $g$ as the function of space such that the stable solitons can also be generated. The idea used in this paper can also be extended to the DNLS equation with other $\PT$-symmetric potentials.

\acknowledgments
The authors would like to thank the referee for the valuable suggestions and comments.
This work was supported by the NSFC under Grant No. 11571346 and and the Youth Innovation Promotion Association CAS.

\end{document}